# Victor J. Emery and recent applications of his ideas

## Serguei  Brazovskii [a,b]

[a] *LPTMS, Bât. 100, Université Paris-Sud, 91405 Orsay, Cedex, France*
[b] *Landau Institute, Moscow, Russia*

**Abstract**

Victor Emery made seminal contributions to the theory of one-dimensional electronic systems and to its applications to organic metals. His inventions became illuminated recently when the joint effect of the ferroelectricity and the charge disproportionation has been discovered in $(TMTTF)_2X$ compounds and beyond. Several of his contributions came to agenda at once: separate gaps in spin/charge channels and the route to solitons, $4k_F$ anomaly, dimerization gap, role of ionic transitions. New phenomena  unify an unusual variety of concepts: ferroelectricity of good conductors, structural instability towards Mott-Hubbard state, Wigner crystallization  in a dense electronic system, ordered $4k_F$ density wave, richness of physics of solitons, interplay of structural and electronic symmetries. The ferroelectric state gives rise to several types of solitons carrying the electron charge, a noninteger charge, spin or both the spin and the charge in special cases. They are clearly observed via conductivity, electric and magnetic susceptibilities. Solitons are challenging for optics where they already seem to determine the pseudogap in absorption. Various features also appear, or are expected, from collective electronic and  coupled electron-phonon modes. The last topic, as well as some aspects of physics of solitons,  recalls also the contributions of M.J. Rice. Moreover, the observation of Mott-Hubbard states refers to classical results of A.A. Ovchinnikov.

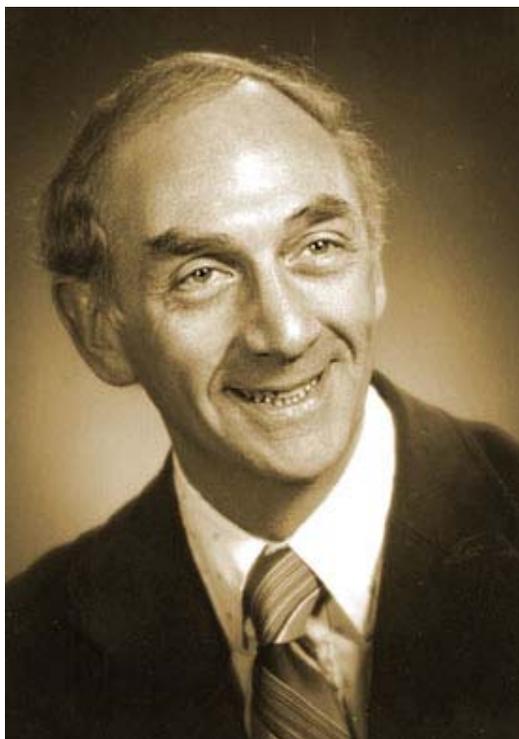

**Victor John Emery: 16 May 1934 – 17 July 2002**

"His experiences with an early digital computer likely persuaded him of the value of more analytical approaches." – *from memoirs about Victor Emery.*

Victor Emery lived as an extraordinary strong personality: humanly, intellectually, physically, and psychologically. He died at the (ever-lasting) summit of accomplishments, withstanding  courageously a devastating illness.

A lone ranger of the APS landscape, Vic Emery got a surprisingly late formal recognition in the US:  the fellow of the American Academy of Arts & Sciences in 2000 and the Oliver E. Buckley Prize in 2001 (with Alan Luther). However, his true authority in science is imprinted in models and solutions bearing his name, in vast citations of his articles and in even wider inspirations among experimentalists and theorists.

For us, Victor Emery is a man of epochs of Organic Metals and of High-Tc superconductors; one of those who made the epochs. Vic Emery came to the science of low dimensional electronic systems both as a curious researcher and as a leading theorist of the exceptional group of Brookhaven National Laboratory. The BNL responsibility, lasting for nearly four decades, was his overwhelming duty, until the very last days of his life. Vic was always close to the neutron scattering experiments at BNL. Thus,  together with J. Axe, he has developed  a theory of the structure factor for 1D crystals in connection to experiments on the "alchemist gold" $Hg_{3-x} AsF_6$ [1] which was showing an intriguing crystalline pseudo long-range order of mercury chains (*cf.* J.-P. Pouget et al at the BNL). For organic metals, Emery (*together with Per Bak: a sad occasion to commemorate another renowned scientist recently passed away*) has proposed a phenomenological theory [2] devoted to the rich phase diagram of the TTF-TCNQ compound. The theory has revealed an unusual interplay of incommensurate structures (cf. R.Comes and G.Shirane, BNL-Orsay) as well as conductance anomalies. The TTF-TCNQ compound was the primary object of worship, in these early days of Synthetic Metals, and still is on agenda today, *cf.* S.Ravy et al.

The most striking feature of the TTF-TCNQ was the so-called $4k_F$ Charge Density Wave - CDW (*cf.* J.P. Pouget et al. in Orsay and also S. Kagoshima in Tokyo) in which the wave number, "$4k_F$", corresponds to one electron per CDW period in comparison with two electrons for conventional, i.e. $2k_F$, CDWs. The theory of Emery [3] assigns an organic metal showing the $4k_F$ anomaly to a detached region of the electronic phase diagram, which is essentially characteristic of strongly correlated systems. Today, we also interpret the $4k_F$ anomaly as the formation of a 1D Wigner crystal thus even building a bridge to semiconducting nano-wires. Recently, the hidden condensation of a commensurate $4k_F$ CDW was identified (*cf.* P.Monceau, F.Nad, S.B., and S.Brown et al) via the new effects of the charge disproportionation and ferroelectricity, which we shall address separately.

A fundamental contribution of that time, also most relevant today, was the work of V.Emery together with A.Luther [4] which results are imprinted in theorists' language as the "Luther-Emery line" and the "Luther-Emery liquid". This work has given a clear view on the spin-charge separation with the direct access to exotic elementary excitations: spinons or holons, described as Sin-Gordon solitons; their classical version was baptized as "φ-particles" by M.J.Rice et al.

Since the discovery (*cf.* D.Jerome et al, Orsay, 1980) of the first organic superconductors, the Bechgaard-Fabre salts $(TMTCF)_2X$, Vic Emery stayed close both to fundamental issues of the theory and to all details on new materials. Thus, he was the first [5] (in parallel to our work with S.Barisic) to pay attention to the effect of a weak dimerization of bonds in $(TMTCF)_2X$ as a drive towards the Mott insulator state. (Here his early work with A.Luther has found the direct application.)

Vic Emery kept being very attentive to new data appearing in the world, sometimes more than people at laboratories of original experiments did. Thus, he realized the importance of tiny structural changes due to subtle transitions of "anionic ordering" AO [5]. (A full rich picture of AOs was published only in 1996, in a review by J.-P.Pouget and S.Ravy written for a memorial volume (*edited by the present author*) of Igor Schegolev, one of patriarchs of Synthetic Metals. Vic Emery offered a statistical model to unify the diversity of structures [6]. He also speculated [7] on a relation between the specific anionic transition in $(TMTSF)_2ClO_4$ and the superconductivity observed in this compound. This question came to agenda recently once again: as a hysteresis at the SDW/SC boundary between the Spin Density Wave and the Superconductivity (*cf.* P.Chaikin et al, D.Jerome et al).

By the mid 80's Vic Emery turned his interests to topics of the Kondo effect, his old passions, where he and co-workers found e.g. solutions for the 1D Kondo lattice.

When the High-Tc epoch broke out, BNL became one of the world leading centers. For Vic Emery that was a time of restless activity, both in his own theoretical work and concerning the impact of his leadership at the BNL and beyond. The first accomplishment of Vic Emery was to suggest a model [8], bearing his name ever since. It describes the $CuO_2$ plane as a system combining features of both the charge transfer and correlation gaps with conducting holes moving over the oxygen network.

The latest passion of Vic Emery became the physics of stripes discovered and studied (*cf.* J.Tranquada et al) at BNL; Vic Emery explored this field with a remarkable activity until his last days. The extravagant view of V.Emery and his co-authors was that stripes might favor the superconductivity, and even bring it to existence in cuprates.

The BNL Bulletin, *v.56, No.27, 9 August 2002*, see also *J.Tranquada et al, Physics Today, v.57, No. 10, p. 92 (2004)*, reviews these last parts of scientific accomplishments of Victor Emery in more details.

## Unwanted thoughts. Dream of integrity.

It is an accident, a very sad one that several prominent contributors, each of them in full intellectual power, have passed away in a short time between the two ICSMs. It could be only a coincidence that all of them were theorists. However, there is something special behind the willing of our community to hold a memorial session devoted to them.

After decades of intense academic activity, Synthetic Metals have reached the focus of material engineering, with wide applications and even commercialization. So it is not quite surprising that the healthy and wealthy pi-conjugated children (polymers, fullerenes, nano-tubes) dominate over their more academic older brothers (CDWs have been pushed away from the cuckoo nest long time ago; organic metals are left for mercy, but the two communities are transparent as reciprocally dark matters, or as participants of parallel sessions at the ICSMs). In theory of conducting polymers and around, analytic approaches have largely been replaced by numerical experiments, usually with a lower accuracy than the real ones. This tendency starts to show up in organic metals as well, while here there are helpful studies on the borderline of analytics and numerics.

Nevertheless, the demonstrated respect to the heritance of prominent theorist uncovers a deep feeling of fundamental grounds of our science and a hidden faith in its integrity. The curious history presented below demonstrates the merits of integrity in an amusing efficiency.

Only the exchange of methods and concepts among CDWs, Organic Metals and Conducting Polymers allowed to resolve the two decades lasting mystery and to turn around a seemingly well-established picture.

## Turmoil of 2000's: ferroelectricity and charge disproportionation in organic conductors.

Plenty of experiments were performed and many theoretical speculations were proposed already within the first years following the discovery of organic superconductivity. The comprehensive picture of the nature of the $(TMTCF)_2X$ family was summarized at the ICSM-1982. Still, the firework of phase was so amusingly rich that it could feed quite intensive researches until our days. Typically, the selenium subfamily $(TMTSF)_2X$ provided the research on the "field induced spin density waves" and the superconductivity while the sulphur subfamily $(TMTTF)_2X$ was most appropriate for effects of the Mott-Hubbard dielectrization. Naturally, the studies were concentrated upon electronic transitions which take place between 1K and 20K, and on fashionable "non Fermi liquid" aspects of the normal state at higher T. Actually, this "normal" part of the phase diagram was filled with tiny structural transitions of diverse "anion orderings" - AO, observed for all "noncentrosymmetric" counterions X (AOs were mostly studied at Orsay: cf. R.Moret, J.-P.Pouget, S.Ravy et al).

Intentionally or not, these "gifts of magicians" were not appreciated neither by experimentalists working on electronic properties, nor by theorists except for some notes due to Vic Emery in 1982 and our (with V.Yakovenko) more systematic approach in 1985. However, the most powerful sleeping bomb was hidden in mysterious "structureless transitions" registered in the sulphur subfamily for centrosymmetric anions X=Br, AsF6, at highest temperatures of typically 150K (cf. C.Coulon et al, 1985). Unexplained and left unattended, these transitions were soon forgotten for the next 15 years. Finally, their time came (cf. F.Nad, P.Monceau, S,B., et al, S.Brown et al, 2001) to overthrow the almost reached consensus on the phase diagram (cf. C.Bourbonnais et al, 1999) and to revise some popular views on the role of electronic correlations (cf. T.Giamarchi at ISCOM-2003). (For detailed history excursions, see S.B. at ISCOM-2001 and more from S.Brown and H.Seo at ISCOM-2003.)

The high temperature region happened to be symmetry broken, moreover in the least expected way: the charge disproportionation (CD) and particularly its striking form of the ferroelectricity (FE) or sometimes Anti-FE. The phenomenon is gaining space by expanding to new compounds (cf. C.Coulon and S.Ravy at ISCOM-2003). In addition, the CD has been known already for a few years among layered organic materials (cf. M.Dressel, H.Mori and H.Seo at ISCOM-2003, H.Fukuyama at ICSM-2002).

In the early 70's, it was already known from I.Dzyaloshinskii and A.Larkin that the special case of chains with one electron per unit cell originates so called Umklapp scattering processes of electrons which drive them to another (Mott-Hubbard) ground state. At the same time A.Ovchinnikov has shown, via analysis of excitation spectra, that this state possesses an optical gap. (For him it was a route to polyenes, where the gap is finally formed by the lattice dimerization (Peierls, rather than Mott-Hubbard, effect) as we know from the success of the SSH model. However, organic metals are generous to give a place for every scenario.) Only the work by A.Luther and V.Emery gave us a clear prescription and handy tools to study this state and calculate its physical properties. To apply it in our case, we introduce the charge phase $\varphi$ as for $2K_F$ CDW/SDW $\sim \cos(\varphi+n\pi/2)$, then the $4K_F$ CDW$\sim \cos(2\varphi+n\pi)$, were the integer $n$ numbers molecules along the stack. We take into account a built-in dimerization of bonds (b) which may (like in TMTCF case) or may not (as in some new compounds) be present, and a spontaneous dimerization of sites (s) which is the charge disproportionation or the condensation of the $4K_F$ CDW. Two types of dimerization give rise to two sources for commensurability and to two contributions $U_s$, $U_b$ to Umklapp interactions:

$$-U_s\cos 2\varphi - U_b\sin 2\varphi = -U\cos(2\varphi-2\alpha), \tan 2\alpha = U_s/U_b.$$

We arrive at a mixed bond/site $4K_F$ CDW with the mixing parameter $\alpha$ which also displaces the mean coordinate $\varphi$ of the whole electronic system, hence the stack becomes electrically polarized. The order parameter $U_s \neq 0$ gives the FE ground state if the same $\alpha$ is chosen for all stacks. The state is the two-sublattice Anti-FE if $\alpha$ signs alternate as in the $(TMTTF)_2SCN$, and more complex patterns have been found in new compounds (cf. S.Ravy at ISCOM-2003).

Since the temperatures $T_0$ of the FE/CD are an order of magnitude higher than the ones of conventional electronic phases, the clearness of their 1D regime becomes uniquely suitable to access the physics of solitons. There are at least two types of solitons in general.

a) For a given $U_s$ the ground state is doubly degenerate between $\varphi=\alpha$ and $\varphi=\alpha+\pi$ which allows for phase $\pi$-solitons: holons with the charge $e$. These carriers have properties of spinless fermions; their pure form is acquired at the so-called Luther-Emery line, which also becomes the line of phase transitions of the CD.

b) Spontaneous $U_s$ itself can change the sign between different domains of the FE displacements. Then the electronic system must also adjust its state from $\alpha$ to $-\alpha$. Hence the FE domain boundary $-U_s \leftrightarrow U_s$ will carry the non-integer charge $q=2\alpha/\pi$ per chain. These $\alpha$- solitons form a gas of quasi-particles at $T>T_0$ and organize themselves in plane domain walls at $T<T_0$ (S.Teber et al). Beyond the above general cases, there are also special presents from the Nature. The most generous one is the subsequent anionic transition of a tetramerization in $(TMTTF)_2ReO_4$ at $T_{AO}<T_0$, which leads to the spin-charge reconfinement. The increase of the conduction gap $\Delta$ and the appearance of a small gap for spin excitations observed at $T<T_{AO}$ signify on a wonderful transformation in nature of solitons, particularly an appearance of special topologically coupled solitons which explore both the charge and the spin sectors.

Within the reduced symmetry, the Hamiltonian becomes

$$H_U = -U\cos(2\varphi-2\alpha) - V\cos(\varphi-\beta)\cos\theta$$

Here θ is the spin phase, such that the gradient $\theta'/\pi$ is the smooth spin density. The new, at $T<T_{AO}$, V- term is the amplitude of the mixed ($\beta\neq0$) site/bond CDW.

Its formation destroys the spin liquid which existed at $T>T_{AO}$ on top of the CD: it opens the spin gap corresponding to creation of the triplet pair of new {$\delta\theta=\pi$, $\delta\phi=0$} purely spin solitons. Next, it prohibits the former $\delta\phi=\pi$ charged solitons, the holons: now they are confined in pairs bound by spin strings. Finally, it allows for combined spin-charge topologically bound solitons {$\delta\phi=\pi$, $\delta\theta=\pi$} which leave the Hamiltonian invariant. For the last composed particle, the quantum numbers are like for the normal electron: the charge e and the spin ½, but their localization is different: the spin is widely extended in compare to charge. Similar effects are expected for any $2k_F$ CDW developed on top of the CD (recall the TMTTF-TCNQ and the Bak-Emery theory). A closer example is the Spin-Peierls state observed in (TMTTF)$_2$PF$_6$ below the CD transition. Hence there is a natural analogy between out treatment of combined solitons and the approach developed by H.Takayama and H.Fukuyama for solitons in the Spin-Peierls state.

The optical gap due to soliton-pair creation is definitely expected to show up in TMTTF compounds but was not seen yet: the whole relevant region is filled by multiple lines of molecular vibrations. Nevertheless, this obstacle is not in vain, being viewed also as another indication for the CD. Indeed, surprisingly (kept noticed since early 80's, cf. C.Jacobsen et al) high intensity of vibrations in TMTTF may be just due to the inversion symmetry lifting by the CD (recall a relevant theory by M.J.Rice). Less pronounced phonon lines in (TMTSF)$_2$X case may tell in favor of a fluctuational regime of the CD; it agrees with a pseudogap (cf. L.Degiorgi, M.Dressel, et al), rather than the true gap, in electronic optical transitions seen, fortunately, below major vibrational peaks. The hidden existence of the CD and the local FE, at least in a fluctuating regime, is the fate of the higher conducting TMTSF subfamily and the major challenge is to detect it.

The ferroelectricity discovered in organic conductors, beyond its own virtues, is the high precision tool to diagnose the onset of the charge disproportionation and the development of its order. The wide temperature-range of the FE anomaly ($T_0\pm30K$) tells us that its development dominates the whole region below and even above these already high $T_0$, and even much higher are the conduction gaps. Remind also the TTF-TCNQ with its ever-present $4k_F$ fluctuations. All that appears at the high energy scale of a "Grand Unification", which knows no differences with respect to interchain couplings, anion orderings, ferro- and antiferroelectric types, between Sulphur and Selenium subfamilies, between old weakly dimerized compounds and the new quarter-filled ones. Hence the formation of the Electronic Crystal (however we call it: disproportionation, ordering, localization or Wigner crystallization of charges; $4k_F$ density wave, etc.) must be the starting point to consider lower phases and the frame to understand their properties. On the theory side, the richness of symmetry-defined effects of the Charge Disproportionation, (Anti)Ferro-Electricity and various

Anion Orderings allows for efficient qualitative assignments and interpretations.

The story of hidden surprises may not be over. Another, interchain, type of the charge disproportionation known in the relaxed (TMTSF)$_2$ClO$_4$ is still waiting for attention, possibly being hiddenly present in other superconducting cases (recall earlier warnings by Vic Emery and by V.Yakovenko and S.B.). Other main challenges for future studies are: hidden existence of CD/FE in the metallic Se subfamily; optical identification of gaps and soft modes; physics of solitons via conductivity, optics, NMR; ferroelectric hysteresis, relaxation, domains.

This rich history tells us about the necessity for reconciliation of different branches of Synthetic Metals. Indeed, the major success in finding the ferroelectric anomaly was due to precise low frequency methods for the complex conduction: designed by F.Nad et al (1993) for pinned CDWs, they were applied "illegitimately" also to SDWs in organic conductors, and finally to structureless transitions. On the theory side, the author's approach of the Combined Mott-Hubbard state has been derived from a fortunate experience (cf. N.Kirova, S.Matveenko and S.B., remind also M.J.Rice and E.Mele) in a model of the "combined Peierls state" in conducting polymers.

The lessons of this surprising story are not limited to organic metals. Hidden or missed effects may be waiting also in pi-conjugated systems: recall (N.Kirova, this session) recent revaluations in theory of optics of polymers which resolve a decade long puzzle of excitons. As a direct link to our content, we notice that recently obtained (T.Matsuda et al) di-substituted polyacetylene is the (AB)$_x$ type polymer which symmetry must provide a net electric polarization, hence either FE or Anti-FE ordering must take place. As a higher caliber smoking gun recall unattended old observation (cf. A.Heeger and M.Winokur) of a perfect columnar orderings of ions in doped polyacetylene (why not to take place in other conducting polymers ?) which must originate a Mott-Hubbard-Emery physics in these systems as well.